\documentclass[aps,twocolumn,preprintnumbers,%
superscriptaddress,floatfix,footinbib]{revtex4}
\usepackage[dvips]{graphics}
\usepackage{amsmath,amsfonts,bm}
\usepackage{epsfig}

\def\be{\begin{eqnarray}}
\def\ee{\end{eqnarray}}
\def\lsim{\mathrel{\rlap{\lower3pt\hbox{\hskip1pt$\sim$}}
     \raise1pt\hbox{$<$}}} 
\def\gsim{\mathrel{\rlap{\lower3pt\hbox{\hskip1pt$\sim$}}
     \raise1pt\hbox{$>$}}} 
\def\la{\langle}
\def\ra{\rangle}

\def\np{Nucl. Phys.}
\def\pr{Phys. Rev.}

\def\la{\langle}\def\ra{\rangle}
\def\del{\partial}
\def\Tr{{\rm Tr}}

\begin{document}
\title{Hidden Local Field Theory and Dileptons\\ in Relativistic Heavy Ion Collisions}

\author{Gerald E. Brown},
\affiliation{Department of Physics and Astronomy,  State University of New York, Stony Brook, NY 11794, USA}
\author{Masayasu Harada},
\affiliation{Department of Physics,  Nagoya University,  Nagoya, 464-8602, Japan}

\author{Jeremy W. Holt},
\affiliation{Department of Physics and Astronomy,  State University of New York, Stony Brook, NY 11794, USA}
\affiliation{Physik-Department, Technische Universit\"at M\"unchen,  D-85747 Garching, Germany}
\author{Mannque Rho},
\affiliation{Institut de Physique Th\'eorique,
CEA Saclay, 91191 Gif-sur-Yvette c\'edex, France}
\affiliation{Department of Physics, Hanyang University, Seoul 133-791, Korea}
\author{Chihiro Sasaki}
\affiliation{Physik-Department, Technische Universit\"at M\"unchen,  D-85747 Garching, Germany}
%
%
%
%
%
\date{\today}
 \begin{abstract}
The notion of ``hadronic freedom'' is introduced based on the vector
manifestation of hidden local symmetry and is used to suggest that the dileptons measured in relativistic
heavy-ion collisions do not provide {\em direct} information on the
spontaneous breaking of chiral symmetry and hence on the mechanism
for mass generation of light-quark hadrons. We give arguments how the dileptons emitted
from those vector mesons whose masses are shifted \`a la Brown-Rho (BR) scaling by the vacuum
change in temperature -- as in heavy-ion collisions -- and/or density -- as in cold compressed matter -- could be strongly suppressed in the hadronic free region between the chiral restoration point and the ``flash
point'' at which the vector mesons recover $\gsim 90\%$ of their free-space on-shell masses and the full strong coupling strength. It may seem ironical that the very mechanism i.e., the vector manifestation in hidden local symmetry, that is to make the mass drop and modify the spectral function is in turn responsible for the dilepton suppression. A possible falsification of this drastic prediction will be indicated. We also briefly discuss the potential role played by the holographic dimension intrinsic in gravity-gauge duality that provides a unified field theory description of hadrons -- both mesons and baryons -- under extreme conditions. Baryons arise as coherent states of pions and vector mesons of holographic QCD which define the ground state or the ``vacuum" on which mesonic excitations could undergo BR scaling.
\end{abstract}

\newcommand\sect[1]{\emph{#1}---}

\maketitle



\section{Introduction}
In QCD, most of the mass of the ground-state light-quark hadrons,
e.g., $\gsim 95\%$ of the $\rho$ meson mass, is understood to be
generated dynamically by the spontaneous breaking of chiral symmetry
in the complex vacuum,
 so it is natural to think that by ``cleansing"
the vacuum, the masses could disappear as the broken symmetry is
restored. Based on this general consideration, a prediction was made
in 1991 that the masses of light-quark hadrons would drop
proportionally to some power of the quark condensate at high
temperature and density~\cite{BR91}. This in-medium property of
hadrons is referred to in the literature as ``Brown-Rho (BR)
scaling."  It was thought that the ``unbreaking" of the symmetry, or
equivalently the ``shedding" of the mass, can be done by heating the
hadronic system to a high temperature (and/or by compressing the
matter to a high density). To unravel the vacuum change induced
specifically by temperature, efforts have been made to look at the
properties of the vector mesons, i.e.\ $\rho$ and $\omega$, in heavy
ion collisions where the temperature could be raised to hundreds of
MeV at which chiral restoration is predicted by QCD to take place.
It was thought that the properties of the $\rho$ (and $\omega$)
meson inside strongly interacting media could be effectively probed
with dilepton pairs, which couple to the vector mesons and provide,
free of final state interactions, a snap-shot of the vector meson
propagating in the medium. Due to the mass shedding in a hot medium,
one naturally expected to observe a downward shift of the invariant
mass of the vector meson in the spectral functions. The measurement
of dileptons, which is the classical tool for discovering new
resonances in particle physics, is used to see how the known $\rho$
resonance in free space moves in a hot medium.

Several such experiments have been performed recently. For reasons
which will become clear below, we will focus specifically on the
NA60~\cite{NA60} experiment, although our discussion applies to
other dilepton measurements such as CERES, etc. What came out of the
experiments so far performed is that no vector mesons affected by the vacuum change
due to temperature and density as predicted by BR scaling are unequivocally 
``seen" by the dileptons. 

In this paper, using a set of arguments, some established and some
conjectural, based on the structure of hidden local symmetry in
low-energy hadronic interactions, we develop the scenario that
dileptons emitted from the vector mesons whose properties are
BR-scaled are strongly, if not completely, suppressed and the
dileptons observed in heavy-ion experiments reflect on-shell
behavior of the vector mesons only modified by the mundane nuclear
many-body interactions~\cite{footnote0} but not by the change of the vacuum due to
temperature and/or density. This means that if our scenario is correct, {\em the BR-scaled vector mesons are almost, if not entirely, ``invisible" to the virtual photon.} As briefly commented in the conclusion section, a similar consideration -- with a small variation -- could apply to dileptons produced in such cold dense environments as expected to be provided at the future FAIR/GSI facility.

We propose as a promising tool to unravel the chiral symmetry structure of hadronic matter a holographic hidden local symmetry framework in which the holographic dimension allows, with a single 5D hidden local symmetry action, a unified description in medium of both mesonic and baryonic excitations treated on the same footing. Baryons emerge here as coherent states of pions and vector mesons which define the ground state or the modified ``vacuum" on which fluctuating vector mesons undergo BR scaling in hot and/or dense medium. A consistent treatment along this line, yet to be made, could lead to a totally different picture from that currently accepted of hot and dense baryonic matter.
\section{The Notion of Hadronic Freedom}
The key proposition in our work is that in a certain range of
temperature (and density~\cite{footnote1}) to be specified below
between the critical $T_c$ at which a chiral transition takes place
and what we will call ``flash temperature" $T_f < T_c$, {\em all}
hadron interactions become weak, as a consequence of which
vector-meson coupling to the virtual photon $\gamma^*$ is
suppressed. This is associated, in a manner to be precisely specified below, with the ``vector manifestation'' (VM)
phenomenon in hidden local symmetry (HLS) developed by Harada
and Yamawaki~\cite{HY:PR}. We first briefly summarize the predictions
of HLS as one approaches $T_c$ from below that is relevant to the
problem at hand and then introduce as clearly as possible the notion
of ``hadronic freedom" that plays the principal role in our discussions.

\subsection{Hidden local symmetry in low-energy strong interactions}
We start with a brief sketch of the 5 dimensional hidden local symmetry action that presumably encodes all low-energy hadron dynamics involving integer spin (0, 1) and half-integer (1/2, 3/2, ...) spin excitations treated on the same footing. One can arrive at an action of the same structure both bottom-up and top-down.

The well-established theory for strong interactions at low
energy is chiral perturbation theory for the
(pseudo-) Nambu-Goldstone pions $U=e^{2i\pi/F_\pi}$, the effective Lagrangian of which has the
form
 \be {\cal L}_{eff}=\frac{F_\pi^2}{4}\Tr (\del_\mu U\del^\mu
U^\dagger)+\cdots\label{chiL}
 \ee
where the ellipsis stands for higher derivative and quark mass
terms. At energies much less than the vector-meson excitation
$m_V\sim 800$ MeV, when treated in a systematic expansion, this represents QCD~\cite{leutwyler}. The
validity of this theory is however highly limited to low energy, so
if one wants to study the vector mesons whose masses are much higher
than that of the pion, chiral perturbation theory with pions alone loses
its power. The question is how to extend the model-independent
approach to higher-energy scale. We will be particularly concerned
with the situation where the vector-meson mass decreases from its
free-space mass $m_V$ to one near the pion mass which is zero in the
chiral limit~\cite{footnote2}. In our approach, this occurs at the
chiral phase transition. This means that we will be dealing with the
possibility -- which will be realized in our model -- that the
vector meson mass becomes comparable to the pion mass and hence both
the vector meson and the pion need to be treated on the same
footing.  This can be most readily done if one exploits local gauge
invariance~\cite{georgi,HY:PR,footnote3}. Seen from the holographic point of view, this is natural as we will argue below.

The astute observation for our development is that the chiral
field~\cite{footnote4} \be U=e^{\frac{2i\pi}{F_\pi}}\in G/H
\ee
where $\pi$ denotes the Nambu-Goldstone (NG) bosons associated with the spontaneous breaking of chiral $G$ symmetry to $H$ with $G=SU(N_f)_L\times SU(N_f)_R$ and $H=SU(N_f)_{V=L+R}$,
has a redundancy when written in terms of the coordinates for the L and R symmetries,
\be
U=e^{\frac{2i\pi}{F_\pi}}=\xi_L^\dagger\xi_R \label{U}\ee
with
\begin{eqnarray}
&& \xi_{\rm L,R} = e^{i\sigma/F_\sigma} e^{\mp i\pi/F_\pi}\ , \
\left[ \, \pi = \pi^a \tau_a/2 \,,\, \sigma = \sigma^a \tau_a/2
\right] \ . \label{def:xiLR}
\end{eqnarray}
The redundancy can be elevated to a local gauge invariance by noting that the chiral Lagrangian built with $U$ is invariant under the local transformation
\be
\xi_{\rm L,R}\rightarrow h(x)\xi_{\rm L,R}
\ee
with $h(x)\in SU(N_f)_V$ and introducing a gauge field $\rho_\mu=\rho_\mu^a\tau^a/2$ that transforms
\be
\rho_\mu\rightarrow h(x)(\rho_\mu-ig\partial_\mu)h^\dagger (x).
\ee
The field $\sigma$ represents the redundancy in the low-energy theory.
Now when written in the lowest order in derivatives in terms of the covariant derivatives $D_\mu\xi_{L,R}=(\del_\mu-ig\rho_\mu)\xi_{L,R}$, the resulting local gauge invariant Lagrangian ${\cal L}^\prime [D_\mu\xi_{L,R}]$ is identical to the chiral Lagrangian (\ref{chiL}) when gauge fixed to unitary gauge corresponding to $\sigma=0$. One says that ${\cal L}^\prime$ is ``gauge-equivalent" to ${\cal L}_{eff}$. To elevate the energy scale from that of the current algebra scale to that of the vector mesons, one adds the kinetic energy term to make the field propagate,
\be
{\cal L}_{HLS}={\cal L}^\prime [D_\mu\xi_{L,R}] -\frac{1}{2}\Tr (\rho_{\mu\nu}\rho^{\mu\nu}) +\cdots\label{HLS}
\ee
The $\sigma$ can be identified with the Nambu-Goldstone (NG) scalars that emerge from the spontaneous breaking of the $H$ symmetry and that are higgsed to give masses to the vector mesons. This is hidden local symmetry theory proposed in \cite{bandoPRL,bandoetal} and later quantized~\cite{HY:PR} that we shall refer to as HLS$_1$.

For our development, it is important that the notion of hidden local symmetry can be extended to an infinite number of hidden local fields. We suggest that HLS$_1$ is a truncated version of a hidden local symmetry theory with an infinite tower of vector fields. As such, HLS$_1$ can be viewed as a realistic model of QCD.

It is clear from the construction of HLS$_1$ that there is nothing
that prevents adding an infinite number of redundancies and
construct a theory with an infinite number of gauge fields. There is
of course the problem of non-uniqueness since the construction would
lead to a large number of possibilities. One such construction which
resembles the nearest neighbor interaction in condensed matter
physics is the construction based on ``theory-space
locality"~\cite{georgi} on lattice which in the continuum limit
leads to a five-dimensional (5D) Yang-Mills action defined in a
curved space~\cite{son-stephanov} \be S_{YM}=\int d^4x dz
\sqrt{g}\frac{1}{2g_5(z)^2}\Tr(F_{AB}F^{AB}) +\cdots\label{5DYM} \ee
with $A=0,1,2,3,z$. The ellipsis stands for higher derivative terms.
Implemented with the Chern-Simons action $S_{CS}$ that encodes
anomalies, this may be considered as a ``dimensionally
deconstructed" QCD~\cite{son-stephanov} emerging bottom-up from the
current algebra Lagrangian. Note that this action contains not only the Goldstone bosons and spin-1 excitations but also baryons as instantons, i.e., coherent states of the Goldstone boson and vector mesons, all on the same footing.

What is most intriguing is that a 5D YM action plus the Chern-Simons
identical in form to the action (\ref{5DYM}) arises also top-down from string
theory. Indeed an action of the form (\ref{5DYM}) was recently
constructed by Sakai and Sugimoto~\cite{sakai-sugimoto} using
D4/D8-D$\bar{8}$ branes valid in the limit that $N_c\rightarrow
\infty$, $\lambda\equiv g_{YM}^2 N_c\rightarrow \infty$ and in the
chiral limit. Now reduced to four dimensions (4D) by the usual
Kaluza-Klein procedure, the action can be phrased in terms of an
infinite tower of vector and axial vector mesons describing
$U(N_f)_L\times U(N_f)_R$ chiral symmetry which may be rewritten in a
local gauge invariant form. We shall refer to  the infinite-tower local
symmetry theory as HLS$_\infty$. Various approximations enter in
arriving at the 5D action such as for instance the ``probe
approximation" $N_f/N_c\ll 1$. But what is robust and generic is that low-energy
hadron dynamics can be encapsulated in a hidden local symmetric action
with an infinite tower of gauge fields. Notice however that
Eq.(\ref{5DYM}) bottom-up is defined in the gauge sector whereas
top-down is in the gravity sector. Generating a gauge invariant theory
from a non-gauge theory as in the bottom-up approach does not lead
to a unique theory and hence can be totally arbitrary unless
constraints are imposed. On the other hand, the top-down approach is
well-defined in the given limit but its UV completion is to a
certain part of string theory which may have nothing to do with QCD.
Furthermore, while the problem we are interested in involves properties
that are intricately locked to the chiral order parameter, i.e., quark condensate, which changes with temperature and density, the large $N_c$ and $\lambda$ limits suppress the
temperature and/or density dependence in the physical variables. One would have to calculate terms higher order in
$1/N_c$. But that would require calculating string loop corrections in the gravity sector which is at present infeasible. For this reason, we are unable to look at vector meson
masses as temperature increases since in the large $N_c$ limit, the
masses do not move. Nonetheless, as we will see below, the infinite tower
structure has an extremely important implication on vector
dominance in hadron electromagnetic form factors and this feature is
expected to play a crucial role in understanding dilepton production
process in heavy ion collisions.

Let us, for the sake of arguments, imagine that we have an HLS$_\infty$ in which all pertinent $1/N_c$
and $1/\lambda$ corrections can be included and which provides a
realistic description of low-energy QCD. Assume further that as
temperature is increased, only the lowest vector meson
$\rho_0=\rho(770)$ drops in mass while the higher lying vector
mesons remain more or less unmodified. This is most likely very difficult
to confirm, but it is plausible. In this case, we could imagine formally integrating out
{\em all} vector mesons except $\rho_0$ and write the resulting
action as a suitable power series in terms of covariant derivatives
of the $\rho_0$ field to get a hidden local symmetric effective
action. How to do this consistently has been proposed by Harada,
Matsuzaki and Yamawaki~\cite{HMY-ads}. What results would be identical in
form to the HLS$_1$ theory of \cite{bandoPRL,bandoetal,HY:PR}. This would justify
HLS$_1$ as a potentially realistic model for low-energy dynamics in
high temperature, particularly if the $\rho_0$ mass does come down to
that comparable to the pion mass which we will argue is what happens
at high temperature in heavy ion collisions. As suggested in
\cite{HY:PR}, the local gauge symmetry plays a crucial role in
treating this situation.
\subsection{The vector manifestation (VM)}
In free space at $T=0$, the vector meson mass $m_{\rho_0}$ can be
taken to be very heavy compared to the pion mass, i.e., $m_{\rho_0}/m_\pi\gg 1$,
particularly in the chiral limit. In this case, there are various different ways
of introducing vector mesons -- with or without local gauge symmetry
-- which are all equivalent at tree order. The hidden local symmetry
approach HLS$_1$ on the contrary takes the vector meson mass to be
of the same order as the pion mass, that is, in the chiral power
counting, \be m_{\rho}\sim m_\pi \sim {\cal O} (p) \ee which is
equivalent to taking the gauge coupling $g$ going as
 \be g\sim {\cal
O} (p).
 \ee
In the framework adopted here where the $\rho$ mass does decrease,
it is very likely that this power counting holds, rendering chiral perturbation expansion justified in the
vicinity of the chiral transition point $T_c$ where the vector meson
mass goes to zero. But it is not obvious why it should hold in free
space where $m_\rho/m_\pi$ is not of order 1. Somewhat surprisingly,
however, it does turn out that chiral perturbation calculation
including the vector meson works fairly well in free space as shown
in \cite{HY:PR}. This indicates that the chiral perturbation which
is justified in large $N_c$ describes the realistic $N_c=3$ world
as an extrapolation.

Given the systematic counting rule,  one can then develop chiral
perturbation. This was done to one-loop order~\cite{footnote5}. In
the parameter space of $g$, $F_\pi$ and $a=(F_\sigma/F_\pi)^2$
(quark masses should of course be additional parameters if we go
beyond the chiral limit.), the renormalization group analysis of HLS$_1$
theory to one loop order reveals a variety of fixed points. This
reflects the well-known non-uniqueness of gauge symmetric theories constructed from non-gauge
theories~\cite{weinberg-gauging}: local gauge symmetric theories can
flow, unconstrained, to a variety of different fixed points. In
order to identify the flow to QCD among them, Harada and Yamawaki
match \`a la Wilson the vector and axial-vector correlators of
HLS$_1$ to those of QCD at a matching scale $\Lambda_M$. The
matching then determines the HLS$_1$ parameters $(g, a, F_\pi)$ in
terms of the QCD variables, namely, $\alpha_c$, $N_c$,
$\la\bar{q}q\ra$, $\la G_{\mu\nu}^2\ra$ etc. that figure in the OPE
of the correlators.

Now when one imposes the condition that at the chiral phase
transition characterized by $\la\bar{q}q\ra=0$,  the vector
correlator $\Pi_V$ equal the axial-vector correlator $\Pi_A$ -- which
is what one expects in QCD, one can identify the fixed point (denoted by bar) that
corresponds to QCD, i.e.,
 \be (\bar{g}, \bar{a})= (0, 1).\label{VM}
  \ee
This point is called the ``vector manifestation" fixed
point~\cite{HY:PR}. Although this fixed point is derived at one-loop
order, it can be seen with a little effort that it is valid to all
orders. The parametric pion decay constant $F_\pi$ does not figure
in the fixed point (\ref{VM}) because there is nothing special about
the flow of $F_\pi$, so at the fixed point, it does not have any
special role as it remains non-vanishing. However the on-shell
(physical) pion decay constant $f_\pi$ which receives loop
corrections $\Delta$ below the on-shell $\rho$ mass,
 \be f_\pi=F_\pi +\Delta
 \ee
should go to zero at $T_c$ because it must track the quark condensate. Note
that this must occur by cancelation. What is important is that as
$T$ goes to $T_c$~\cite{HS-T} (and as density $n$ goes to the
critical density $n_c$~\cite{HKR}), the hadronic matter flows to the
VM fixed point (\ref{VM}),
 \be
g^*&\sim& \la\bar{q}q\ra^*\rightarrow 0,\nonumber\\
(a^*-1)&\sim& {\la\bar{q}q\ra^*}^2\rightarrow 0.
 \ee
Here and in what follows the asterisk stands for temperature (or density) dependence.
This means that the
parameters of the Lagrangian have an intrinsic dependence on
temperature locked to the property of the condensate
$\la\bar{q}q\ra$. This has the key consequence that the {\em parametric
mass} of the vector meson very near the VM fixed point behaves
as~\cite{footnoteM}
 \be
{M^*_\rho}^2=a^*{F^*_\pi}^2 {^*}{g^*}^2\sim {\la\bar{q}q\ra^*}^2,\label{para-mass}
 \ee
so will go to zero. This is the {\em statement of BR scaling} sharpened by
HLS$_1$.

Now what about the physical (or pole) mass of the vector meson
$m_\rho$ in medium? For this one has to compute thermal loop terms
which will necessarily involve the gauge coupling, so
 \be {m^*_\rho}^2=a^*
{g^*}^2 {F^*_\pi}^2 +{g^*}^2 B \label{pole-mass}
 \ee
where B is a smooth function of temperature
given by thermal loop corrections. As $g^*\rightarrow 0$, both terms
go to zero as $\sim {g^*}^2$, so that BR scaling should still hold for
the physical mass. It is however to be noted that far away from the
VM fixed point, there is nothing that suggests that the physical
mass must be dropping according to BR scaling. In fact, it can go up
or down or even remain more or less unchanged depending on the thermal loop
corrections involving complicated many-body nuclear dynamics. In
order to ``see" BR scaling in terms of HLS$_1$ in temperatures away from the VM fixed point, therefore, one needs to
bring in baryons via topology in addition to the bosonic degrees of freedom that are present explicitly. Such a calculation remains to be done. The same consideration holds in cold dense matter as discussed in the conclusion section.

\subsection{The hadronic freedom}
In the close vicinity of the VM fixed point, the interactions
are likely to be weak since the gauge coupling goes to zero.
The main reason for this is that in HLS$_1$, the strongest
interaction at low energy which is in the p-wave channel, i.e., the
$\rho$ channel, is suppressed by the vanishing $g$. Further, the
s-wave $\pi\pi$ interactions are governed by derivative interactions
(in the chiral limit), so should be weak at low energy. There may be
other degrees of freedom not explicitly taken into account in
HLS$_1$ -- such as quasiquarks or possible scalars -- but the
quasiquark interactions mediated by the vector mesons are also
suppressed by the gauge coupling and quasiquark-pion interactions
are derivative-coupled, so suppressed by the power counting. Thus it
seems reasonable to take {\em all} hadronic interactions to be weak
near the VM fixed point. Now how far down in temperature can one
ignore the interactions? In principle, the intrinsic temperature
dependence will be running as temperature is varied. For instance,
the gauge coupling in the chiral limit will scale up from zero at
$T_c$ to near its free space value at some temperature which we
shall call ``flash temperature" $T_f$.

\subsubsection*{Flash temperature $T_f$}
The flash temperature $T_{f}$ can be defined following the work
of Shuryak and Brown~\cite{shuryak-brown} of the STAR data~\cite{star-data}. Shuryak and Brown determined the flash temperature to be the temperature at
which the $\rho$ was 90\% on-shell, namely, when the $\rho$ mass is
700 MeV. Whereas the flash temperature is independent of the
centrality, this is not the case for the freezeout temperature
$T_{freezeout}$. In fact $T_{freezeout}$ decreases as the centrality
increases.  Thus, given $T_{freezeout} \simeq T_{f}$ for
peripheral collisions, $T_{freezeout} < T_{f}$ for central
collisions and during the time the system is between these two
temperatures the pions from $\rho$ and $a_1$ decay will be
rescattered as it is no longer possible from their detection to work
backward to the parent vector mesons.

It is possible to pin down the numerical value of the flash
temperature from lattice calculations. We shall extract this
quantity from Miller's lattice calculation~\cite{Miller,footnoteL} which we
analyzed in \cite{BLR05}. The result is
 \be
T_{f}\approx 120\ {\rm MeV}.
 \ee
How this result comes about can be summarized as follows.   As
pointed out in \cite{BLR05}, in Miller's lattice calculation of the
gluon condensate, the soft glue starts to melt at $T\approx 120$
MeV. The melting of the soft glue, which breaks scale invariance as
well as chiral invariance $dynamically$ and is responsible for
Brown-Rho scaling, is completed by $T_c$ at which the particles have
gone massless according to the VM. One can see that the gluon
condensate at $T\sim 1.4\ T_c$ is as high as that at $T\sim T_c$.
This represents the hard glue (or ``epoxy") which
breaks scale invariance $explicitly$ but has no effect on the hadron
mass. We see that the melting of the soft glue is roughly linear,
implying that the meson masses drop linearly with temperature.~\cite{footnoteL}

How the (bare) parameters of the Lagrangian scale {\em intrinsically} in temperature (and/or density) is not quantitatively known except in the very close vicinity of the VM fixed point. In the absence of guidance, theoretical or experimental,  we take, as the first step, the simplest scenario which is to ignore interactions {\em entirely} between the chiral restoration
point $T_c$ and the flash point $T_f$ and let the relevant massless
degrees of freedom coming down from above $T_c$~\cite{parketal} flow
without interactions as the system cools to $T_f$. We then assume that hadrons go
(nearly free-space) on-shell at $T_f$ recovering their (nearly) full
interaction strength and free-space mass. Corrections due to increasing temperature as one approaches $T_f$ -- involving both the intrinsic dependence and the thermal loops -- need of course to be computed at the next stage of refinement. This region between $T_c$ and $T_f$ will be dubbed ``hadronic freedom regime."~\cite{footnoteMH}

We must admit that the notion of hadronic freedom is a strong
assumption which needs to be confirmed. At present, there are no
theoretical tools to check this notion in unambiguous ways. As pointed out below, it could
be validated or refuted by experiments. All we can
say at the moment is that while there are indirect evidences for it,
there are no ``smoking gun" evidences against the notion of hadronic
freedom either. What we propose to do in this article is to {\em posit} the
hadronic freedom and then see how far we can go in confronting
nature. We will then present our interpretation of its consequence
on dileptons in heavy ion collisions. Our scenario, which is drastically
different from the conventional scenario, has the merit of providing a possible, albeit indirect,
link between the manifestation of BR scaling in low-energy nuclear structure physics -- some of which will be mentioned in the conclusion section -- and possibly direct manifestation or precursor effects in heavy ion processes.

\subsection{Vector dominance and infinite tower}\label{subsec:VDIT}
In considering dileptons in heavy ion collisions,  it is customarily
assumed that vector dominance holds in medium at {\em all} temperatures
and densities. In HLS$_1$, vector dominance does hold in free
space but not in hot/dense medium. In fact, vector dominance is violated at $T=T_c$, with a drastic effect on dileptons in hot/dense medium.
We briefly review how this comes about and then
discuss what happens in the presence of the infinite tower of vector
mesons figuring in holographic QCD that follows from gravity/gauge duality
in string theory. We believe that the infinite tower structure encoded in the holographic direction which treats both mesons (spin 0 and 1) and baryons on the same footing, when $1/N_c$ corrections are suitably handled,
could ultimately clarify the behavior of hadrons in high temperature and/or density.

\begin{figure}[ht]
 \centerline{\epsfig{file=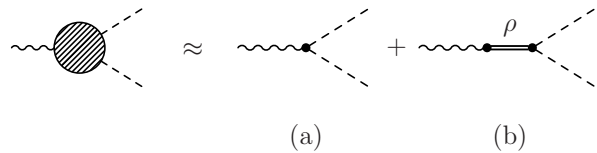,width=8.5cm,angle=0}}
\caption{Leading contributions to the electromagnetic form factor
  of the pion. (a) direct $\gamma\pi\pi$ and
(b) $\gamma\pi\pi$ mediated by $\rho$-meson exchange. In terms of
HLS$_1$ Lagrangian, (a) is proportional to $(1-a/2)$ and (b) is
proportional to $a/2$.
  }
 \label{VD1}
\end{figure}
In HLS$_1$ theory, the photon couples to the charged pion as in
Fig.\ref{VD1}:  (a) direct $\gamma\pi\pi$ with the coefficient $(1-a/2)$
and (b) $\gamma\pi\pi$ mediated by $\rho$-meson exchange with the
coefficient $a/2$.
It turns out that $a=2$ in medium-free ($T=n=0$) space which gives the
KSFR relation correctly.
This value completely suppresses the direct
photon coupling Fig.\ref{VD1}(a), giving full strength to
Fig.\ref{VD1}(b). This is the famous lowest vector-meason dominance (LVD for short)
of the pion form factor, i.e. $\rho$-meson dominance in Sakurai's
sense~\cite{Sakurai}.
Note however that $a=2$ is not on the trajectories of the
renormalization group equation for $a$, so it is in
some sense an accident~\cite{HY:VD}. Thus unsurprisingly, this LVD is violated -- in fact ``maximally'' -- as
$T\rightarrow T_c$: As $a$ approaches 1, the direct coupling
enters with the same strength as the vector exchange
term~\cite{HS-T,HS:VD}.
This violation of LVD is argued below to be
largely, if not wholly,
responsible for the suppression of dileptons in the hadronic
freedom regime.

An interesting possibility is that we might be able to understand the violation
of LVD in HLS$_1$ in terms of the infinite tower in HLS$_\infty$. In the gravity/gauge
dual model of Sakai and Sugimoto (SS)~\cite{sakai-sugimoto} which has the
correct chiral symmetry structure of QCD, the pion form factor is
vector dominated by the infinite tower:
 \be
F(q^2)=\sum_{n=0}^\infty\frac{g_{\rho_n}g_{\rho_n\pi\pi}}{m_{\rho_n}^2-q^2}
=\sum_{n=0}^\infty\xi_n\frac{g_{\rho_n\pi\pi}}{1-q^2/m_{\rho_n}^2}\label{FF}
\ee where \be \xi_n=\frac{g_{\rho_n}}{m_{\rho_n}^2},
 \ee
and $g_{\rho_n}$ and $g_{\rho_n\pi\pi}$ represent
respectively the $\gamma$-$\rho_n$ coupling and the $\rho_n$-$\pi\pi$
coupling. These couplings are determined in the
large $N_c$ and $\lambda$ limit by the equations of motion for the
$\rho_n$ fields integrated over $z$ (the fifth coordinate) which
represents energy spread and are fixed once $N_c=3$, $f_\pi$ and the
Kaluza-Klein (KK) mass $M_{KK}$ are taken from the meson spectra.

At the photon point $q^2=0$,
one has the charge sum rule for the charged pion,
 \be
F(0)=1=\sum_{n=0}^\infty\xi_n
g_{\rho_n\pi\pi}=\bar{\xi}\sum_{n=0}^\infty
g_{\rho_n\pi\pi}\label{sumrule}
 \ee
where we have used the observation~\cite{sakai-sugimoto,HRYY} that \be
\xi_n=\frac{g_{\rho_n}}{m_{\rho_n}^2}=0.272\pm 0.001\equiv \bar{\xi}
\ee which holds for $n=0,1,2,3$. We have no proof but we will assume
that it holds for {\em all} $n$. A surprising result obtained by
Sakai and Sugimoto is that the sum rule (\ref{sumrule}) is saturated
by the lowest four $\rho$'s, i.e., $\rho_n$ with
$n=0,1,2,3$,
\be
F(0)=1.31-0.35+0.05-0.01=1.00.\label{sr} \ee
There is a tantalizing empirical evidence that this sum rule is satisfied in $\tau$ decay~\cite{tau}. We will return to this observation later in connection with multi-pion-photon coupling.

It should be noticed that the formula in Eq.(\ref{FF})
is obtained by using the equations of motion for the vector meson fields.
Thus the sum rule Eq.(\ref{sr}) holds only when the couplings have no appreciable momentum dependence.
Furthermore, $1/N_c$ corrections which are expected to be important in the time-like region may not be negligible here. Actually, as shown in Fig.~2 of Ref.~\cite{HS:dilepton},
the experimental data of the pion form factor
are well fit in the time-like region by the one obtained in the HLS$_1$ with only the $\rho_{n=0}$ contribution,
in which $a\simeq2$ is used and about 10\% momentum
dependence is added to $g_\rho$
together with an appropriate $1/N_c$ correction to the width.


To understand the potential role of the higher members of the tower in the form factor, it is convenient to separate the form factor (\ref{FF}) into ``low" and ``high" components as
\be
F(q^2)=F^{low} (q^2) +F^{high} (q^2)\label{s}
\ee
with
\be
F^{low} (q^2)&=& \sum_{n=0}^k\frac{g_{\rho_n}g_{\rho_n\pi\pi}}{m_{\rho_n}^2-q^2},\label{low}\\
F^{high} (q^2)&=&
\sum_{n=k+1}^\infty\frac{g_{\rho_n}g_{\rho_n\pi\pi}}{m_{\rho_n}^2-q^2}.
\label{high} \ee The sum rule (\ref{sr}) suggests that we take
$k=3$. Now for low $|q^2| \lsim m_{\rho (770)}^2$, we can drop $q^2$
in $F^{high}$ (\ref{high}) and find from (\ref{sr})
 \be
F^{high} (q^2)\approx
\sum_{n=3}^\infty\frac{g_{\rho_n}g_{\rho_n\pi\pi}}{m_{\rho_n}^2}
\approx 0.\label{zerocharge}
 \ee

Our conjecture is that the statement $a\rightarrow 1$ as
$T\rightarrow T_c$ is tantamount to the statement that in HLS$_\infty$ the higher
tower members for $n=4,..., \infty$ in the large $N_c$ limit
and those for $n=1,..., \infty$ with $1/N_c$ corrections included
figure ``maximally".
How the infinite tower figures in the
nucleon EM form factor is illuminating in this connection. As
discussed in \cite{HRYY}, the VD in the nucleon form factor is
violated in HLS$_1$. In fact, it is described rather well by $a\sim
1$ in Fig.\ref{VD1} with the pion replaced by the nucleon field.
In HLS$_\infty$, the higher members of the tower are found to play the
role of the point-like coupling of HLS$_1$~\cite{HRYY}.

Unfortunately the conjecture cannot, at present, be verified in the
SS model. In HLS$_1$, it is the vanishing of the quark condensate as
$(a-1)\propto \la\bar{q}q\ra^2\rightarrow 0$ that ``kills" the point
coupling. However it is known that the quark condensate has no
temperature dependence in the large $N_c$ limit. This suggests that
to verify the conjecture, $1/N_c$ corrections are needed:
Although a part of the $1/N_c$ corrections could be included by
hadronic loop corrections in the gauge sector~\cite{HMY-ads},
certain $1/N_c$ corrections in the gravity sector -- so far inaccessible -- may give important contributions.
\section{Some consequences that follow from the hadronic freedom}
In this section, we address how our approach based on the hadronic
freedom in HLS$_1$ fares with experiments in heavy ion collisions.
Our attitude here will be as follows. We posit the hadronic freedom
together with BR scaling that we infer from HLS$_1$ as the premise
of our approach and discuss the implications on a variety of
processes that can be accessed by the approach. We will first treat
the cases where our approach can make clear statements and then present
what can be said about the dileptons in heavy ion collisions that
have been controversial in the literature such as the NA60 data etc.
\subsection{The STAR $\rho^0/\pi^-$ ratio}
As a case where the notion of the hadronic freedom makes a simple prediction, we recall how the STAR $\rho^0/\pi^-$ ratio (STAR ratio for short)~\cite{star} can be understood.

In a recent experiment, STAR collaboration has reconstructed the $\rho$-mesons from the two-pion decay products
in the Au + Au peripheral collisions at $\sqrt{s_{NN}}= 200$ GeV. The ratio found was
\be
\frac{\rho^0}{\pi^-}|_{STAR}=0.169\pm 0.003 ({\rm stat})\pm 0.037
({\rm syst}),\label{stardata}
 \ee
almost as large as the $\rho^0/\pi^-=0.183\pm 0.001 ({\rm stat})\pm
0.027 ({\rm syst})$ in proton-proton scattering. The near equality
of these ratios was not expected, in that the $\rho$ meson width
$\Gamma\sim 150$ MeV in free space is the strongest meson
re-scattering that one has. For instance, at SPS one needs about 10
generations in order to get the dileptons~\cite{rapp}. Furthermore,
if one assumes equilibrium at the freezeout, then the ratio is
expected to come to several orders of magnitude smaller, say,
$\frac{\rho^0}{\pi^-}\sim 4\times 10^{-4}$~\cite{peteretal}.

There may be a variety of explanations for this observation, some of
which are along the standard line of thought from which our scenario
appears to depart drastically. What distinguishes the result that
follows from the hadronic freedom -- which turns out to come out consistent with
the experiment (\ref{stardata}) --  is that the mechanism is
extremely simple if one assumes that the $\rho^0$'s are
reconstructed at the flash point at which the $\rho$ meson emerges
from the hadronic freedom regime going through only one generation
from $T_c$. We should however stress that no claim is made here that
the proposed mechanism is either unique or the only viable
explanation. It is just that our scenario has the merit to be simple and consistent
within the proposed framework anchored on the hadronic freedom and the vector
manifestation of HLS$_1$.

This matter was discussed in \cite{BLR-STAR} to which we refer for details. Here we review it briefly for the sake of making our arguments self-contained.

The key point in our scenario is that because of the decreased width
due to the hadronic freedom, the $\rho$ meson is assumed to go
through {\it only one generation} before it freezes out at the flash
point in the peripheral collisions in STAR. There will be no
equilibrium at the end of the first generation.

As the plasma decreases in temperature towards $T_c$ from above,
lattice gauge calculations\cite{asakawaetal} show a set of 32
$SU(4)$ degenerate vibrations\cite{footnote6}: scalar, pseudoscalar,
vector and axial vector, i.e., the set of degrees of freedom that
figure in Weinberg's mended symmetry at the chiral transition
point~\cite{weinberg}. The widths become large as $T\rightarrow T_c$
from above.~\cite{footnoteW} We connect these with the $SU(4)$ group of
mesons, including the $\rho$ of the HLS$_1$ vector manifestation,
becoming massless (except for a small spin-spin interaction effect)
as $T$ moves up to $T_c$ from below. Aside from the background
$\eta$ and $\omega$, these 32 mesons decay as they go on shell to
give 65 pions (three of them being already present in the 32 $SU(4)$
mesons)~\cite{BLR,BGR}. Thus, taking the 22 $\pi^-$ mesons and
subtracting 3, which are traced back to the vertex inside the plasma
where the $\rho^0$ was formed, one has
\begin{equation}
\frac{\rho^0}{\pi^-}\approx \frac{3}{22-3}\approx 0.16
\end{equation}
in agreement with the experiment. As noted, this number is about the same as
the $\rho^0/\pi^-$ ratio measured in $pp$ scattering. However, the
vacuum $\rho$ has a width of $150$ MeV so its lifetime is only
$\hbar/150 \simeq 4/3$ fm/c and in proceeding from $T_c$ to the
flash point where the $\rho$ goes on shell, which we shall find to
be at $\sim 120$ MeV, were the interactions unsuppressed, there
would be several generations of $\rho \rightarrow 2\pi \rightarrow
\rho$, whereas our estimates show that there are none; the $\rho$'s
move from $T_c$ to $T_{f}$ without interaction. Thus had the $\rho$
possessed its on-shell mass and width with $\sim$ five generations
as in the standard scenario, the $\frac{\rho^0}{\pi^-}$ ratio would
be closer to the much smaller equilibrium value.
\subsection{Dense matter}
There are certain consequences of the hadronic freedom that are more
prominent in dense matter than in hot matter. Since dilepton
experiments involve density in addition to high temperature, some of
them need to be explained.

One of the most important applications of the notion of the hadronic
freedom is to the role of kaon condensation in compact stars. A
recent publication addresses this matter in conjunction with
black-hole formation and a possible cosmological
consequence~\cite{kaon-BLR}.

While the phase structure of dense matter is not well understood,
the first important phase change that can take place in dense matter
above the nuclear matter density $n_0$ and that can be treated reliably thanks to the VM is kaon condensation at a
density $\sim 3n_0$. This phase transition is considered to be
responsible for the ultimate fate of compact stars, e.g., the formation of stable neutron stars or collapse into black holes~\cite{kaon-BLR}. It takes
place because of the decrease of the kaon mass and the increase of the
electron chemical potential at increasing matter density: Matter density drives
the mass of the kaon downwards and the electron chemical potential upwards so that they cross
in the hadronic freedom region between
$n_c$ and the flash chemical potential $n_f$. The
dropping gauge coupling~\cite{HKR} and the rapid approach to
$a=1$ in dense medium~\cite{MR-halfskyrmion} enable one to reliably
calculate the mass of the kaon as a function of density in the
vicinity of the VM fixed point and hence determine the critical
density. This phenomenon has a far-reaching consequence on the physics of compact stars. {\em What is important for our problem at hand is that when matter density is present, the vector dominance is more rapidly violated than in temperature alone}. This is highly relevant in our picture in NA60.

\section{Dileptons}
We now apply the hadronic freedom picture to dileptons produced in
relativistic heavy ion collisions. Our conclusion will be that if
the hadronic freedom picture is correct, then the dileptons produced
from the $\rho$ mesons populating the interval between $T_c$ and
$T_f$ are strongly suppressed and hence the measured dileptons do
not provide a direct snapshot of the properties of the vector mesons
that are BR-scaled.
\subsection{The spectral function near the VM fixed point}
Consider first the thermal $\rho$ spectral function in HLS$_1$
theory in the close vicinity of the VM fixed point approached from below. We shall show
that depending upon whether the $\rho$ width is vanishing or
non-vanishing, one gets a completely different result for the
spectral function $\Re$.

Suppose that we have a renormalized $\rho$ propagator in the form
\begin{equation}
D_\rho(q^2) =
 \frac{- 1}{ q^2 - {m_\rho^\ast}^2 + i \sqrt{q^2} \Gamma_\rho^\ast
 }\label{prop}
\end{equation}
with
\begin{equation}
{m_\rho^\ast}^2 = {M_\rho^\ast}^2 + R \ .
\end{equation}
where $M_\rho^\ast$ is the mass parameter with intrinsic temperature
dependence in the HLS$_1$ Lagrangian given by (\ref{para-mass}), $R$ is the real part of the
self-energy and $\Gamma_\rho^\ast$ is the width coming from the
imaginary part of the self-energy. The asterisk stands for an in-medium
quantity. To make the discussion more general than
HLS$_1$, we will suppose that the self-energy term could contain
contributions from degrees of freedom other than just $\rho$ and
$\pi$. For instance, $\Gamma_\rho^\ast$ could in principle contain
many-body effects such as collisional width, involving baryonic
excitations. The spectral function relevant for the dilepton
spectrum at a fixed $T$ (including $T > T_{f}$) is expressed
by the imaginary part of the vector current correlator as
\begin{equation}
{\Re}\equiv {\rm Im} G_V(q^2=s)
 = {\rm Im}
  \left[
    F_\rho^2(s) D_\rho(s)
      \right]
\  \label{Im Gv}
\end{equation}
where $F_\rho (s)$ is the $\rho$ decay constant which in general
will have an imaginary part. Ignoring the imaginary part, we have
\begin{equation}
F_\rho^2(s) = {f_\sigma^\ast}^2 \
\end{equation}
where $f_\sigma^\ast$ is the in-medium decay constant of the
longitudinal component of $\rho$ (see Eq.(\ref{def:xiLR})). Note
that ${f_\sigma^\ast}^2$ includes both elementary hadronic and
thermal loop corrections. It is related to the in-medium $\rho$-$\gamma$
coupling $g_\rho^\ast$ as
\begin{equation}
{g_\rho^\ast}^2 = {m_\rho^\ast}^2 {f_\sigma^\ast}^2 \ .
\end{equation}

Setting $s = {m_\rho^\ast}^2$ at the in-medium pole position, we get
for the spectral function
\begin{eqnarray}
\Re (s={m_\rho^\ast}^2) =
 \frac{ {g_\rho^\ast}^2 }{ {m_\rho^\ast}^3 \Gamma_\rho^\ast }
 \ . \label{Im Gv 2}
\end{eqnarray}
In terms of the leptonic decay width which is given by
\begin{equation}
\Gamma_{ee}^\ast = \frac{4\pi\alpha^2}{3}
  \frac{ {g_\rho^\ast}^2 }{ {m_\rho^\ast}^3 }\ ,
\end{equation}
we have
\begin{eqnarray}
{\Re}(s= {m_\rho^\ast}^2) &\sim&
 \frac{ \Gamma_{ee}^\ast }{ \Gamma_\rho^\ast }
\ ,
\end{eqnarray}
where we have neglected certain numerical factors such as
$\frac{4\pi\alpha^2}{3}$. This shows that the behavior of the
spectral function depends on the properties of both the leptonic and
hadronic widths.

Let us see what happens when one approaches the chiral transition point at which $g^*=
f^*_\pi=f^*_\sigma=0$. For this, we look at Eq. (\ref{Im Gv 2}):
 \be
\Re (s={m_\rho^\ast}^2)\propto \frac{ {f_\sigma^\ast}^2 }{ {g^\ast}
\Gamma_\rho^\ast }
 \ee
where we have used $m_\rho^*\sim g^*$ near the fixed point. We need
to know (a) how $g^*$ and $f_\sigma^\ast$ approach the fixed point
and (b) more crucially the behavior near the fixed point of the
hadronic width $\Gamma_\rho^\ast$. In HLS$_1$, it is established
that $g^*\propto \la\bar{q}q\ra\rightarrow 0$ but how $f^*_\pi$ (or
$f^*_\sigma$) behaves in terms of the quark condensate is not known.
It is because it goes to zero at the fixed point by the cancelation
between the intrinsic term $F_\pi^*$ and thermal loop corrections,
both of which are not zero.

Suppose for the sake of argument that the in-medium pion decay
constant goes as $f^*_\pi\sim g^*$. Then the behavior of the
spectral function will be entirely dependent on whether the hadronic
width vanishes or not. If it is non-vanishing, then the spectral
function will go to zero. However if it vanishes, how it vanishes
will matter. Within HLS$_1$, we expect ${ \Gamma_\rho^\ast }\sim
{g*}^2$. In this case the spectral function will be singular. This
would be the case if leptons decoupled from the $\rho$ {\em after}
hadrons (e.g., pions) did. If this case holds, then the quantitative estimate made on the suppression factor discussed in section \ref{seen} would have to be taken with caution.

In reality, the situation is not at all clear. First of all, there
is the threshold mass required by the lepton mass (as in NA60) which
prevents the approach to the fixed point. Furthermore there can be
several elements or mechanisms that are not encoded in HLS$_1$ at
the perturbative level which can intervene to modify the picture. We
have already mentioned, among other things, the possible role of the
infinite tower implied by holographic dual QCD, the solitonic
degrees of freedom (e.g., collisional broadening), Hagedorn
excitations etc. This means that we cannot simply take the HLS$_1$ result mentioned above at its face value. Now given our ignorance of the variety of mechanisms that could possibly invalidate the simple HLS$_1$ prediction, we choose to accept the experimental indication, described in section \ref{seen}, that {\em no} dileptons are observed from the $\rho$ mesons living in the temperature interval between $T_c$ and $T_f$ that we are identifying with hadronic freedom.
It would of course be extremely interesting if future experiments did exhibit a sharp peak at a low dilepton invariant mass. It would provide an evidence for the validity of the simplest picture of HLS$_1$ and of the vector manifestation fixed point. It would however falsify the role of the assumed hadronic freedom on the strong suppression of the dilepton coupling to the vector mesons in the given temperature (and density) range.
\subsection{Suppression of dileptons in the hadronic freedom}
Although similar arguments apply to other dilepton experiments (such as CERES, PHENIX etc.), we
will focus here on the conditions that are met in NA60.

There are several sources in NA60 for the dimuon suppression in the
hadronic free region in temperature from $T_c$ to $T_{f}$ as well as
in density from $n_c$ to $n_f$. In this region, the $\rho$ meson
decay to a lepton pair is suppressed both in the EM sector and in
the hadronic sector. Notable among the possible suppression
mechanisms are:
\begin{enumerate}
\item {\bf VD violation}:  In the EM sector, the maximal violation of VD discussed above reduces
the $\rho\rightarrow  l^+ l^-$ decay rate by a factor of 4. This
reduction sets in more quickly in the presence of baryonic matter as
mentioned in section IIIC.
\item {\bf Wave function suppression:} At the VM fixed point, the transverse
$\rho$ decouples from the vector current as the gauge coupling $g$
goes to zero. If we were to consider in terms of the quasiquark
picture, this would imply, in the hadronic sector, the suppression
of the bound zero-mass quasiquark-quasiantiquark wave function of
the $\rho$ meson, $|\Psi_\rho (0)|^2$.  This effect is encoded in
HLS$_1$ theory by the suppression of the $\gamma\rho$ coupling which
gives a factor of $\sim 2$ reduction.
\item{\bf Phase space suppression:} Because of the dropping gauge
coupling $g$, $a\approx 1$ and the dimuon threshold mass of 210 MeV,
the phase space for the decay is suppressed by a factor $\sim 2$.
This effect is not entirely orthogonal to the wave function suppression
effect as explained below.
\end{enumerate}
These mechanisms will give a suppression factor relative to the standard picture~\cite{footnote-F} of the
dileptons of $\sim 10$ in the hadronic freedom regime where the BR
scaling effect is to be operative in HLS$_1$ theory.

We elaborate on these points.

\subsubsection{VD violation}
The violation of VD comes about by $a^*$ going to 1 in hot/dense
medium. Although one should perform the renormalization group
calculation to find $a^*$ in terms of temperature as well as density
-- such a calculation has not been done yet, there is a strong indication
that $a^*$ goes rapidly from 2 to 1 in hot/dense medium. One can see in
the work of Shuryak and Brown~\cite{shuryak-brown} that this drop is
sudden, at $T\approx 120$ MeV, with increasing $T$. In the presence of
density, it should drop even faster as argued in section IIIC.
Therefore, we assign $a^* = 1$ to the region from $\sim 120$ MeV to $\sim 175$ MeV.
This gives the suppression factor of about 4. Moreover, there will be dileptons coming from the direct $\gamma \pi \pi$ coupling that have nothing to do with the in-medium properties of the $\rho$ meson. These should contribute a roughly constant background to the total dilepton yield. Note that the range of
$\rho$-meson energies is 0 to 700 MeV, so that we are discussing
$\rho$-mesons with energy below the on-shell $\rho$-mass of 770 MeV.
Therefore, we should look to the low mass part of the spectrum.

\subsubsection{$\rho$ wave function suppression}
There will be further hindrance for the dileptons which come through
the $\rho$ meson on the hadronic side, because the dileptons are
similar to a photon, to the extent that their masses are small
compared with their invariant energy, so that there is an
approximate factor
\begin{equation}
F=|\psi_\rho(0)|^2
\end{equation}
(where $\psi_\rho$ is the $\rho$ wavefunction)
of the probability of the off-shell $\rho$ meson being at its
origin. This gives rise roughly to a factor of 2 reduction. This can
be seen in HLS$_1$ as follows: The photon couples to the $\rho$ with
the coupling $\sim a^*g^*{F_\pi^*}^2$ which goes to zero when the VM is approached,
i.e., $g^*\rightarrow 0$. The vanishing of the photon coupling
manifested in the vanishing gauge coupling represents the vanishing
of the $\rho$ wave function at the origin, $|\psi_\rho(0)|_{m_\rho^*
=0}=0$, when the mass goes to zero. What is relevant to us is the
pion-loop medium correction $\delta$ to the $\gamma\rho$ coupling
$g_{\gamma\rho}$
 \be
 \frac{g^\ast_{\gamma\rho}}{g_{\gamma\rho}} \sim \frac{a^\ast \,
g^\ast (1+\delta)}{a g }
 \ee
with $F_\pi^*/F_\pi\approx 1$. The density dependence of $\delta$
has not yet been computed in HLS$_1$ but its temperature dependence
has been computed perturbatively in Ref.~\cite{HS:dilepton}:
 \be
\delta(T=T_{f}) &\approx& -0.2, \nonumber\\
\delta(T=T_c) &\approx& -0.4 \label{perturbative}
 \ee
for $T_{f}\approx 120$ MeV and $T_c\approx 175$ MeV. This increases
the reduction factor by an additional factor $\sim 1.6-2.8$. Given
that the density effect is not included in the loop correction, it
seems reasonable to take the resulting reduction factor to be $\sim
2$.

We have to stress here that the estimates (\ref{perturbative}) were
made in perturbation theory with HLS$_1$ Lagrangian with $\rho$ and
$\pi$ only, so may not be reliable in the vicinity of the phase
transition point. There may be non-perturbative effects which
require going beyond the chiral perturbative approach to HLS theory.
Furthermore, the drastically simplified HLS structure with the
ground state vector mesons only may be inadequate in describing
dynamics near the phase transition. As mentioned above, such degrees
of freedom as the infinite tower of vector mesons encoded in
holographic QCD, baryonic excitations via solitons (i.e., skyrmion
or instanton) etc. might play a crucial role.

\subsubsection{Phase space suppression}
One can arrive at a similar reduction factor from the phase
space consideration which seems to have a partial overlap with the wave function
suppression considered above. Integrating the contribution from potentially BR-scaling
$\rho$ mesons down from $T_c$, we first find that the integral
begins at $2m_\mu \sim 210$ MeV, because the muon masses must be
furnished, and then continues down to the nearly on-shell $\rho$
mass of $\sim 700$ MeV, the remaining 10\% to put the $\rho$
completely on shell coming from the kinetic
energy~\cite{shuryak-brown}. However, in our scenario, the $\rho$ masses are
essentially zero~\cite{footnote7} at just above $T_c$
\cite{parketal} behaving as if they have a quark substructure of two
massless quarks, coupled to a massless $\rho$. In the integral going
down from $T_c$ the integrand, beginning from zero at $2m_\mu$,
would be expected to increase as the $\rho$ goes back towards the
value of the on-shell $\rho$ because of the effect on
$\bar{\psi}_\rho \psi_\rho(0)$ from the increasing correlations as
the $\rho$ goes on shell. It seems reasonable to approximate the
entire effect of $\bar{\psi}_\rho \psi_\rho(0)$ by a linearly
increasing function which starts from zero at $2m_\mu$ and goes to 1
at $T=T_{\rm flash}=120$ MeV, where the $\rho$ mesons go on shell.
This will cut the dileptons that go through the $\rho$ spectral
function down by a factor of $\sim 2$.

\section{Dileptons ``seen" in the experiment}\label{seen}
We have argued that the dileptons produced from the $\rho$ mesons
between $T_c\approx 175$ MeV and $T_f\approx 120$ MeV are strongly
suppressed. The question then is: What are the dileptons observed in
NA60 (and also in CERES)? We suggest that the answer to this is that
they are almost, if not entirely, {\em produced outside} of the hadronic
freedom region, namely, at the flash temperature $T_f$.  We present
arguments to support this suggestion.

\subsection{Dileptons in the vicinity of the $\rho (770)$}
\begin{figure}[htb]
   \begin{center}
\epsfig{file=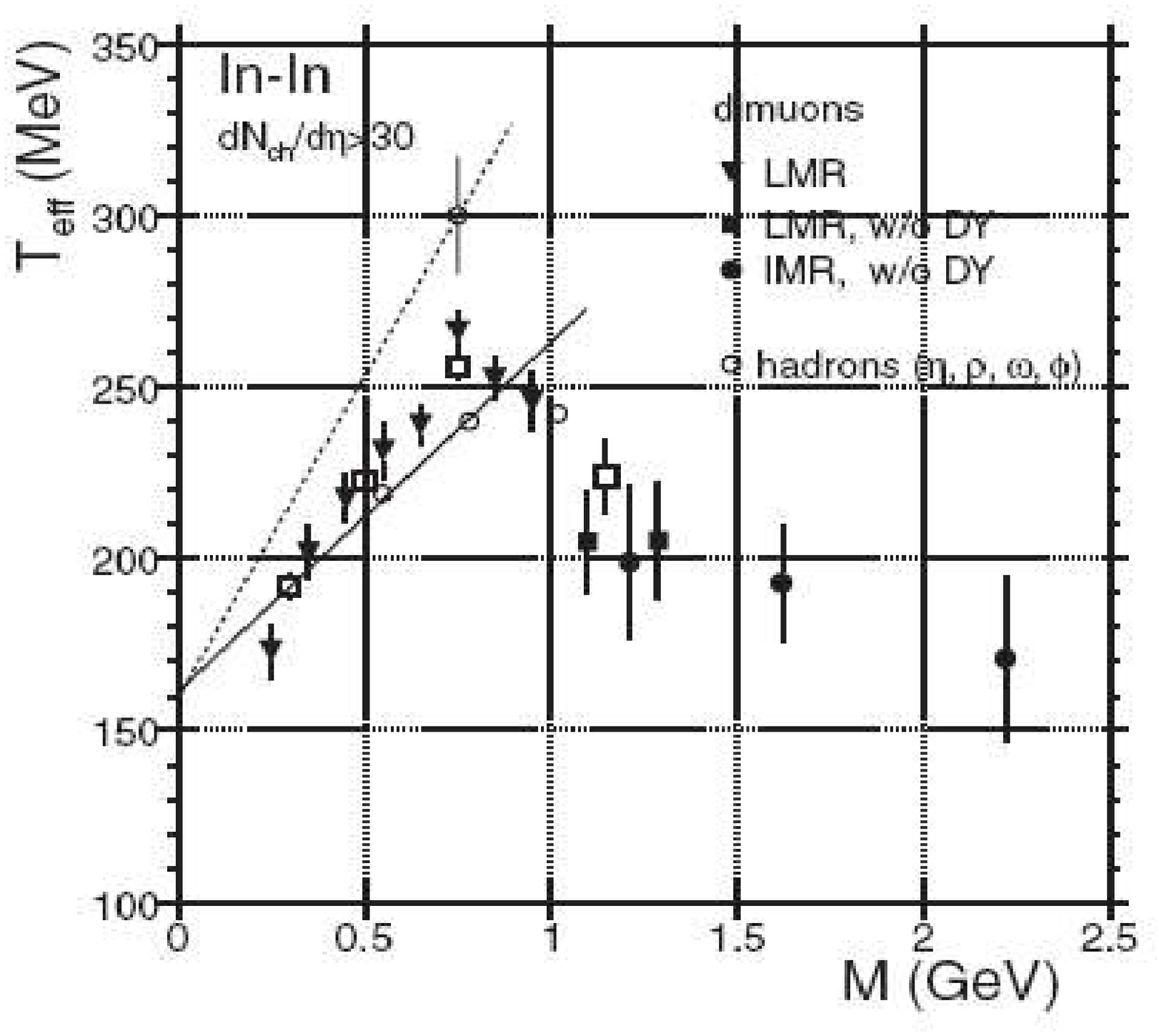, width=7.5cm} \vskip -0.5cm \caption{Out of
the blue-shifted hadrons, the dileptons from the $\rho$ are shifted
the most. They are reconstructed on the (steepest) dashed line.
Between the dashed and solid lines, the $\omega$, $\phi$ and $\eta$
will flow somewhat less than the $\rho$'s, but nonetheless are blue
shifted. The $T_{\rm eff}$ for the $\rho$ has $300 \pm 17$ MeV for
the peak, and $231 \pm 7$ MeV for the underlying continuum in the
window $0.6 < M < 0.9$ MeV. Reproduced from Fig.\ 4 of
\cite{arnaldi}.}
     \label{arnaldifig}
   \end{center}
\end{figure}

For our purpose, we consider most useful the article
on radial flow of thermal dileptons by R.\ Arnaldi {\it et
al.}~\cite{arnaldi}. In Fig.\ref{arnaldifig} is reproduced their
plot of the $T_{\rm eff}$ obtained in an effective theory along the
line of flow, which produces a blue-shifting
\begin{equation}
T_{\rm eff} = \sqrt{\frac{1+\beta}{1-\beta}}T
\end{equation}
and in which the Boltzmann factor $m$ is replaced by $m_\perp$, the
mass perpendicular to the flow. Here $\beta$ refers to the flow.
Since the flow develops only late, after the hadrons have gone on
shell, the amount of blue shift tells one about how late the hadrons
emerge~\cite{footnote8}.
By disentangling the peak from the continuum Arnaldi {\it et al.}
find $T_{\rm eff} = 300 \pm 17$ MeV for the peak and $231 \pm 7$ MeV
for the underlying continuum in the window $0.6 < M_{\mu\mu} < 0.9$
MeV. The latter is suggested as coming from the $\eta$, $\omega$,
and $\phi$, which freeze out earlier due to their smaller coupling
to the pions, the latter producing the flow. Note that in our scenario, the $\omega$
and $\eta$ are included in the $SU(4)$ of particles which go
massless at $T_c$.

\begin{figure}[htb]
   \begin{center}
\vskip 0.5cm
 \epsfig{file=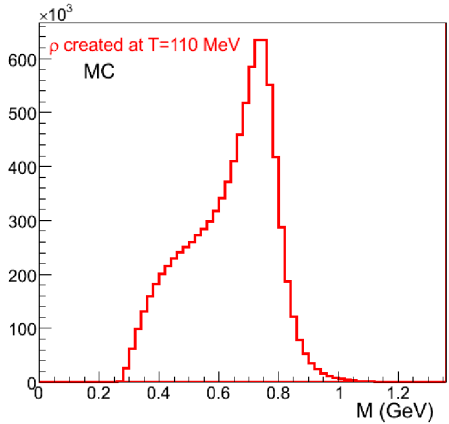, width=7.5cm}
\vskip -0.5cm
 \caption{The on-shell
$\rho$ reconstructed by NA60 assuming $T_{\rm freezeout} = 110$ MeV
and $m_\perp$ scaling across the $\rho$ resonance (not very
sensitive). Reproduced from \cite{dam}.}
     \label{dam}
   \end{center}
\end{figure}

The $\rho$-meson that has been reconstructed as the peak on the
broad continuum~\cite{dam} is shown in Fig.~\ref{dam}.
We can interpret this as the $\rho$ meson that emerges from the hadronic free region and goes on-shell at the
flash point. To confirm the validity of this interpretation, let the peak height at the on-shell
mass $m_\rho=770$ MeV -- which fixes the normalization -- be denoted
by $D$ which we take to be $D\approx 620 \times 10^3$ from Fig.~\ref{dam}. We take the quantity given by  Fig.~\ref{dam} to be of the form
 \be {\rm Fig.~\ref{dam}}=D F(M_{\mu\mu})
\left(\frac{m_\rho}{M_{\mu\mu}}\right)^2\label{figure}
 \ee
where $M_{\mu\mu}$ is the dimuon invariant mass and $F$ is the
relativistic form of the Breit-Wigner formula,
 \be
F(M_{\mu\mu})=\left(1+\left\{\frac{1-(M_{\mu\mu}/m_\rho)^2}{\Gamma_\rho/m_\rho}\right\}^2\right)^{-1}.\label{BW}
 \ee
Let us see whether Fig.~\ref{dam} coincides with or deviates appreciably from the spectral function of the on-shell $\rho$ with the free-space $\rho$ width $\Gamma_\rho=150$ MeV.  This can be easily checked. Taking $M_{\mu\mu}=$ 400, 600 and 780 MeV in Eq.~\ref{BW}, we find, in units of $10^3$, 153, 202 and 593 respectively. These are essentially what's given by Fig.~\ref{dam}, i.e., 160, 260 and 600.
Thus modulo the normalization, the $\rho^0$'s in the flow are all on-shell $\rho$'s {\em unaffected} by the medium. There is evidence for neither BR-scaling $\rho^0$'s nor $\rho^0$'s with broadened widths.

So far we have not considered explicitly what $a_1$ mesons can do. Dusling, Teany and Zahed~\cite{dusling} fit well the part of the dilepton spectrum below the $\rho$ meson mass around $M=500$ MeV by invoking effects from the $a_1$ meson at 1260 MeV. In our scenario, the $a_1$ meson makes up part of the 32 degrees of freedom, massless at $T_c$, that are flowing without interactions from $T_c$ down to $T_f$. At $T_f$, the $a_1$, having an immense on-shell width of $\sim 200$-$650$ MeV, suddenly goes on shell, changing into a $\rho+\pi$. This $\rho$ joins the flow. This contribution which doubles the effective number of $\rho^0$'s at $T_f$ is included in our scenario in the normalization of Fig.~\ref{dam}. The $a_1$ effect of Dusling et al. could however figure in our scenario outside of the hadronic freedom regime and outside of the flow as mentioned below.
\subsection{Dileptons below the $\rho (770)$ peak}
%
Let us now turn to how one can understand the dileptons that contribute to the
spectral function {\em below} the free-space $\rho$ mass. The principal
thrust of this paper was that the bump near $\sim 770$ MeV is
populated predominantly by near-on-shell $\rho$'s that are separated
by the blue-shifting in the effective thermodynamics flow, and that
in the invariant mass region below $\sim$ 770 MeV in which BR
scaling vector mesons should be present, the photon coupling is
highly suppressed in accordance with the hadronic freedom. As a
result, those dileptons produced by mundane mechanisms, i.e.,
``background," should dominate the spectral function, masking more
or less completely BR-scaled vector mesons. We suggest that the
most likely mechanism that produces the dominant component of the
``background" is pions in heat bath in strong interactions that populate at or just below the
flash point with the Shuryak-Brown kinematical effects taken into account.

Our basic premise is that in the hadronic freedom scenario, all
hadrons, including pions, flow freely from the critical temperature
to the flash point in a time of $\gsim 3$ fm/c. (As mentioned, pions  will also be
weakly interacting like other hadrons, because the hidden gauge
coupling $g$ -- which governs the p-wave $\pi\pi$ interaction
through $\rho$ exchange -- is small.)  In this regime, the photon
has a direct coupling to a pion due to the violation of vector
dominance with $a\sim 1$ -- which would reflect the presence of the vector manifestation --
but relatively few leptons are expected to be produced by the pions in that regime.

Reaching the flash point, however, all hadrons making up $\sim 32$
degrees of freedom going on-shell, with the hidden gauge coupling
regaining its free-space strength, will produce dileptons with the
vector dominance recovered with $a$ going to 2. The vector mesons
produced in the interactions at the flash point will go into flow and appear at
on-shell mass.

Now in the central collisions, the freezeout temperature is lower than
the flash temperature, so the hadrons which go on shell at the flash
temperature go several generations before they freeze out at
$T_{freezeout}$. The large number of pions present between $T_f$ and
$T_{freezeout}$ -- not in the flow -- will interact strongly
producing, among others, copious $\rho^0$'s which will emit the lepton
pairs. These $\rho^0$'s, interacting strongly in medium, will develop widths that
are due to conventional nuclear interactions -- including
``sobars"~\cite{BR-NA60} -- in the form of the $\rho$
self-energy. These will populate the invariant masses below the $\rho
(770)$ peak. The $a_1$'s so produced could also emit dileptons \`a la
Dusling et al. We suggest that most, if not all, of what's calculated in the
references~\cite{others,dusling} belong to this class of contributions
which have nothing to do with what's happening in the hadronic freedom
regime.

\begin{figure}[hbt]
\includegraphics[width=6cm, angle=0]{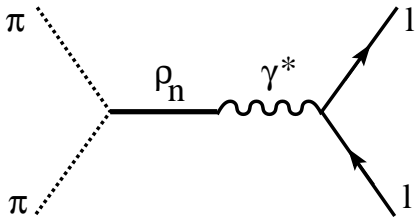}
\caption{Dilepton ($l=e,\mu$) production by $\pi^+\pi^-$ annihilation in heavy ion collisions involving the infinite tower of vector mesons $\rho_n$ $n=0,1,2,..., \infty$} \label{diag1}
\end{figure}

If one considers the infinite-tower structure of HLS$_\infty$, there
can be additional sources for dileptons of lower invariant mass. Let
us consider how the higher members of the tower can contribute to the
dilepton production. Let us first look at their contribution in the
process $\pi^+\pi^-\rightarrow l^+l^-$ ($l=\mu, e)$. The relevant
graph is given by Fig.~\ref{diag1}.
This is given by the form factor (\ref{FF}) separated into two components as (\ref{s}). Here we are dealing with the time-like regime for which $1/N_c$ corrections need to be included. Let us imagine that they can be calculated. Then they will give both real and imaginary parts. The real part will shift the masses and the imaginary part will give the widths. We expect that the low-mass component (\ref{low}) will essentially give the Breit-Wigner form
\be
F(M^2_{l^+l^-})\approx
\left(1+\left\{\frac{1-(M_{\mu^+\mu^-}/m_{\rho (770)})^2}{\Gamma_\rho/m_{\rho (770)}}\right\}^2\right)^{-1}
\ee
with the free-space $\rho$ width $\Gamma_\rho=150$ MeV. As for $F^{high}$, we can ignore the self-energy correction and $|q^2|=M_{\mu^+\mu^-}$ which are expected to be much smaller than the higher-tower vector-meson masses in (\ref{high}) and find that due to the charge sum rule, Eq.~(\ref{zerocharge}) applies to the form factor in the time-like regime. It may be possible to justify this argument with the time-like form factor obtained from the $\tau$ decay measurement mentioned above~\cite{tau}. Thus essentially the high tower does not affect the structure that is obtained in HLS$_1$ with $a=2$ in the time-like regime as in the space-like regime.

What we have shown above is that the 2$\pi$ contributions to dileptons are constrained by the pion charge sum rule involving the infinite tower of vector mesons in hQCD. However there are no such constraints for multi-pion contributions. Specifically in HLS$_\infty$, $2n\pi$ for $n>1$ can contribute importantly via higher $\rho_n$'s. Of course there is the threshold effect so that the $4\pi$ contribution will be lower-bound by $\sim 560$ MeV. In fact there is indication that $4\pi\rightarrow e^+e^-$ receives substantial contributions from $\rho (1450)$ and $\rho (1700)$. This indication comes from a recent analysis of the available CMD-2 and BaBar data implementing the three lowest members of the tower and $a_1$ mesons~\cite{4pi}. The complete vector dominance structure of hQCD is not present in the analysis of \cite{4pi}, so it is not obvious what the infinite tower structure will do in heavy ion processes. We expect that such four-pion contributions mediated by higher members, if present, could make non-negligible contributions to dileptons with invariant masses $\gsim 560$ MeV.

\section{Comments and Conclusions}
Brown-Rho (BR) scaling was a prediction based on effective field theory,
not a hypothesis or a conjecture as often stated, in a
phenomenological attempt to build the main feature of chiral
invariance, the behavior of the broken chiral symmetry which gives
hadrons, other than the pion, masses. A great deal of activity had
been spent in the years preceding BR scaling to introduce
quarks as substructure of the hadrons into nuclear physics. Whereas
there were some successes, such as in the chiral bag model, color
transparency, etc., there was not a broad theoretical basis for the
many-body problem. A local field theory that did just that by
matching to QCD was introduced by Harada and Yamawaki in HLS$_1$
which gave the first effective field theory support for BR
scaling. BR scaling also came out naturally in
strong-coupling lattice gauge calculations in the large $1/d$
expansion (where $d$ is the number of space
dimension)~\cite{ohnishi}.

The original idea of BR scaling was that in medium, because
of background density or temperature which affects the ``vacuum" structure,
hadron masses, other than that of the pion, that are dynamically generated had to be scaled as a function of the quark condensate. In many-body systems where such a vacuum change is generated by strong nuclear interactions, this idea can be addressed only in an effective field theory framework and not directly in QCD language. So the question then is: In what way would BR scaling manifest in complex many-body processes? The answer to this question depends on scales that are involved. This point was recognized in the early phenomenological papers by Brown and Rho where the scaling effect was looked for in spin-isospin dependent nuclear interactions, notably the tensor forces. Thus very near the chiral transition point, HLS$_1$ theory, well-defined by the VM fixed point, makes an extremely simple and clear-cut prediction -- whether right or wrong -- in the property of the vector mesons $\rho$ and $\omega$ and hence of physical observables associated with them. However away from the VM fixed point, the situation is quite different and a lot more intricate. While the parameters of the HLS$_1$ Lagrangian are directly linked to the quark condensate thanks both to the local gauge invariance and to the matching to QCD, most physical observables, while dependent on BR scaling, have complicated dependence on the chiral order parameter. For instance, at the scale of the structure of light nuclei, BR scaling can figure in giving effective nuclear forces. Most striking of all in this category is the archaeologically long lifetime of the $^{14}$C decay explained in terms of the tensor force that results from an intricate cancelation between the pionic term and the $\rho$ term where BR scaling enters through the parametric mass term~\cite{c14}. On the other hand, near the Fermi liquid fixed point of nuclear matter, BR scaling manifests itself in the Landau Fermi liquid parameter $F_1$, explaining how exchange currents affect the orbital gyromagnetic ratio in heavy nuclei~\cite{friman-rho}. It is in this context (as stressed in \cite{BR-NA60}) that one would have to understand the role of dileptons as a possible litmus indicator for chiral symmetry in hot and/or dense matter.

Using the notion of the hadronic freedom, we have argued that the dileptons measured in heavy ion collisions are mostly, if not all, from the $\rho^0$'s outside of the hadronic freedom (or fireball) regime. We should stress that the hadronic freedom, a notion derived from HLS$_1$ with its VM fixed point, has an additional assumption extraneous to HLS$_1$ in the VM that requires verification. It assumes that the $\rho$ width at the VM fixed point is not zero whereas a strict adherence to HLS$_1$ with no additional ingredients -- which we believe is not complete -- would imply that the width vanishes. This renders it to a possible falsification. An observation of a strong enhancement of the dilepton production in the {\em close vicinity} of $T_c$, while giving a partial support to the simplest HLS$_1$, would falsify the notion of the hadronic freedom and hence our scenario on dileptons described in this paper.

The violation of vector dominance, a characteristic of HLS$_1$ in the vector manifestation, could also be subject to falsification. While dileptons from $\rho$ mesons are suppressed, one expects dileptons coming from pions via direct coupling due to the VD violation. To see this effect, it would be necessary to subtract {\em all} the ``background" coming from on-shell sources. This might be a daunting task but if such effect is found, it will be a ``smoking-gun" signal for HLS$_1$ in the vector manifestation and also an indirect support for BR scaling.

There have been discussions, both experimental and theoretical, on the behavior of vector mesons in cold dense nuclear medium~\cite{footnoteCS}.
Experiments in the upcoming accelerators will surely generate a lot of activities in this area. It would be difficult to probe directly the density regime in the close vicinity of the VM fixed point, so what one can do is to study various precursor effects of matter at densities near that of nuclear matter. The JLab~\cite{jlab}, CD-Bonn~\cite{omega} and KEK~\cite{KEK-PS} experiments belong to this class.

We now check what is involved in HLS$_1$ theory for such effects.

What is addressed in this class of experiments is the ``mass" of the vector mesons $\rho$ and $\omega$ ``seen" inside nuclear medium at a density $\lsim n_0$. The measured quantities reflect the pole mass which in HLS$_1$ is given by
what is equivalent to Eq.~(\ref{pole-mass}) in density,
\be
{m_V^*}^2=a^*{F_\pi^*}^2{g^*}^2 + \Sigma^*\label{dense}
\ee
where the asterisk here stands for the {\em intrinsic} dependence on the chiral order parameter, i.e., the quark condensate, which changes with density and $\Sigma^*$ is the ``dense loop" corrections gotten from the HLS$_1$ Lagrangian in a suitable perturbative series. Now to address the dense loop corrections, it is necessary to introduce fermionic degrees of freedom to implement matter density. In principle this could be done in terms of dense skyrmions in HLS$_1$ or better in HLS$_\infty$ involving the infinite tower of vector mesons in 4D or instantons in 5D which automatically takes into account the intricate vacuum structure in which both pions and vector mesons are condensed generating a self-consistent baryonic background for fluctuating mesons. In a sophisticated version of this type of approach, one could also build in $N^*$'s, thereby incorporating such collective excitations as ``sobars"~\cite{BR-NA60} etc. Efforts are being made to arrive at dense matter through this approach but up to date there has been little progress along this line.

What has been done up to date is to introduce, somewhat ad hoc, quasiquarks in a gauge invariant way into HLS$_1$~\cite{HKR}. At one loop order, we can rewrite (\ref{dense}) as
\be
{m_V^*}^2=a^*{f_\pi^*}^2{g^*}^2 + 12 {g^*}^2 cG^*\label{dense1}
\ee
where $f_\pi^*$ is the pion decay constant renormalized by dense one-loop corrections, $G^*$ is a known function of density and $c$ is the only undetermined constant in the Lagrangian which is of ${\cal O}(1)$. We can make a rough estimate of (\ref{dense1}) at nuclear matter density~\cite{footnoteF},
\be
(m_V^*/m_V)^2(n_0) &\approx& (f_\pi^*(n_0)/f_\pi)^2 +0.21 c\nonumber \\
&\approx& 0.64 + 0.21 c\label{dense2}
\ee
where we have used the value $f_\pi^*/f_\pi\approx 0.8$ quoted in the literature at nuclear matter density~\cite{kienle-yamazaki}. If $c>0$ and $c\sim {\cal O}(1)$, one sees that the dense loop corrections -- that comprise essentially of mundane nuclear interactions in the given field theory framework -- are comparable to the first term that purports to provide direct information on BR scaling.  We see from this simple calculation that in order to reveal unequivocally the role of BR scaling, one first has to define precisely what quantities one is to zero-in on in a precisely defined effective field theory, and then to compute many-body corrections in a way consistent with the effective theory so defined. Phenomenological Lagrangians constructed solely to fit free-space quantities -- hence valid at tree order and devoid of the Wilsonian renormalization-group flow~\cite{HY:PR} -- are not likely, if not impossible, to provide useful information of this subtle nature. In HLS$_1$, it is the parametric mass (first term on the RHS of (\ref{dense1})) -- and not the measured pole mass -- that carries the looked-for information.

That dileptons, considered to be an ideal snapshot of the chiral order parameter, appear to be blind to BR scaling may seem to indicate that BR scaling cannot be probed unambiguously. This is however not the case. There are in fact certain quantities that could provide a clear-cut signal for the BR scaling at work. One such case that has been studied is the pion velocity $v_\pi$ at $T_c$. The $\rho$ mass becoming degenerate with the pion mass as $T\rightarrow T_c$ in HLS$_1$ in the VM makes the pion velocity rapidly approach 1~\cite{SUS-qq}, whereas in the absence of the massless $\rho$, $v_\pi$ can be rigorously shown to vanish~\cite{son-stephanov-vpi}. This is a night-and-day difference. The pion velocity going to zero would be a strong indication that the $\rho$ mass does not drop, hence the vector manifestation is not realized in nature and would call for a revamping of the hidden gauge structure in the form of HLS$_1$, perhaps requiring the infinite tower of vector mesons. It would be a challenge to both theorists and experimentalist to find other cases where HLS$_1$ in the vector manifestation can be exposed equally directly and ultimately to exploit HLS$_\infty$ in exploring the intricate nature of matter near the critical point. In this connection, it is of importance to note that in holographic QCD, baryons emerge as coherent states of pions and vector mesons, in particular, the $\rho$ meson, and the vector modes ``seen" in medium are excitations on top of this ground state in which both pions and vector mesons are condensed. Thus the property of vector mesons in baryonic medium is expected to be vastly more intricate than the naive picture would give.

\subsection*{Acknowledgments}
Three of us (MH, MR and CS) acknowledge the hospitality of Yukawa
Institute of Theoretical Physics where ``Yukawa International
Program for Quark-Hadron Sciences" entitled {\it ``New Frontiers in
QCD 2008"} was held and where a part of this paper was written. The
work of MH was supported in part by
the JSPS Grant-in-Aid for Scientific Research (c) 20540262
and the Global COE Program
``Quest for Fundamental Principles in the Universe''
of Nagoya University provided by Japan Society for the
Promotion of Science (G07),  part of the work of MR was funded by the WCU project of the Korean Ministry of Educational Science and Technology and
CS acknowledges partial support by DFG
cluster of excellence ``Origin and Structure of the Universe''.


\begin{thebibliography}{99}
\bibitem{footnote0} By ``mundane" nuclear processes, we mean those processes involving hadrons, both mesons and baryons, in nuclear medium that are calculable starting with a phenomenological chiral Lagrangian -- possessing pertinent symmetries -- whose parameters are fixed in the matter-free vacuum. By definition, calculations are strictly valid at tree order although in practice loops are commonly computed with phenomenological form factors or cut-offs.
\bibitem{footnote1} Unless otherwise stated, a similar argument applies to density although we will address only
temperature.
\bibitem{footnote2} In nature, the quark mass is not zero, so we are
not at the chiral limit but barring some special cases, it is a good
approximation. We will be considering the chiral limit unless
otherwise stated.
\bibitem{footnote3} Going systematically to higher energy scales with gauge-fixed
Lagrangians is known to be very difficult if not impossible. It is
the local gauge invariance that allows a systematic account of
higher derivative terms and renders the effective theory amenable to
a UV-completion.
\bibitem{footnote4} Here $N_f$ is for any number of flavors but
will be taken to be $N_f=2$ for applications.
\bibitem{footnote5} Going beyond one-loop order is highly
daunting and at present is unpracticable not only because there are a
large number of diagrams that enter but also a large number of
parameters to be fixed, which is not feasible given the present
state of experimental data. It should however be mentioned that the
results given below are valid at least qualitatively to all orders of chiral expansion.
\bibitem{footnoteM} The parameteric mass will be denoted as $M_\rho$ while the physical (pole) mass will be given as
$m_\rho$.
\bibitem{footnoteMH}
An interesting possiblity to explore is that the existence of the infinite tower of mesons in the hadronic freedom regime could figure in cooling of the system from $T_c$ to $T_f$.
\bibitem{footnote6} As explained in detail in \cite{BLR-STAR}, we are taking into account scalar and
axial-vector degrees of freedom in the HLS framework although their
inclusion has not been rigorously formulated.
\bibitem{footnoteW}  This suggests yet another feature that
departs drastically from the standard scenario that depicted a
strongly-correlated hadronic matter below $T_c$ making transition to
a non-interacting quark gluon plasma above $T_c$. Our scenario
instead has it that a non-interacting nearly massless hadronic
matter below $T_c$ makes a transition to a strongly-correlated
matter just above $T_c$.
\bibitem{footnote-F} By ``standard picture" we mean the descritpion in which vector dominance is assumed and
mundane many-body effects are taken into account in both the real and imaginary part of self-energy of the vector meson in the temperature interval involved. It may or may not incorporate some sort of ``dropping mass" effects. As noted above, this factor cannot be applied without suitable modifications to the HLS$_1$ result of \cite{HS:dilepton}.
\bibitem{footnote7} The $SU(4)$ symmetry at just above $T_c$ is slightly broken
 by the spin-dependent force, but the inertial parameter $p_0$ is large, so the mass is only a few percent of the
free $\rho$ mass.
\bibitem{footnote8} We are not concerned with the sudden decline (and therefore flow) in
$T_{\rm eff}$ at masses $>1$ GeV which is most naturally explained
as a transition to partonic processes like $\bar{q}q \rightarrow
\mu^+ \mu^-$.
\bibitem{footnoteCS}
Finite 3-momenta carried by $\rho$ mesons are expected to affect
in-medium $\rho$ spectral functions and hence obscure the manifestation of chiral symmetry in nuclear medim. The calculation performed in hadronic many-body theory for the $\rho$ produced by photon
reactions off nuclei~\cite{riek:jlab} has indeed shown that the relatively
large 3-momenta relevant for the JLAB CLAS experiments significantly affect the pole structure of the propagator in medium and hence the spectral function of the $\rho$ meson. Thus, the dilepton rates with
a low momentum are more relevant to details of medium effects in the data.
\bibitem{footnoteF} Let us rewrite Eq.~(\ref{dense2}) in a form that would be given in the standard (phenomenological) approaches with no local gauge symmetry and without matching to QCD -- hence no intrinsic density dependence -- as $(m_V^*/m_V)^2=1+\Delta$. It is the $\Delta$ that the standard approaches would interpret -- and the experimentalists measure -- as the ``mass shift." In our simple calculation with HLS$_1$, it is given by $\Delta=0.12c-0.36$. We see that for $c>0$ and $c\sim {\cal O} (1)$, the ``mass shift" can be negligible due to cancelation, the extent to which the cancelation takes place depending upon the detailed structure of the nuclei studied. In this form, BR scaling is not ``visible." This illustrates the difficulty in probing BR scaling at a density far from the VM fixed point.
\end{thebibliography}
\end{document}